# The Early Career Astrobiology Workforce Under Strain: Survey Evidence from 2025-2026

Dzurilla, K. A.[1*], Rizzo, G.[2] Johnson, P.E.[3], Monreal, P. J.[4] Elavarasan, I.[5], Spiers, E. M.*[6,7]

[1]Scientific Society for Astrobiology; [2]School of Biological Sciences, University of Nebraska-Lincoln; [3]Purdue University; [4]School of Oceanography and Astrobiology Program, University of Washington; [5]University of Texas Rio Grande Valley, [6]University of Texas at Austin, Institute for Geophysics; [7]Woods Hole Oceanographic Institution

**Abstract:** A survey of self-identified early career astrobiologists was organized by the NOW early career group FLOW and the Scientific Society of Astrobiology. During November 2025–February 2026, 165 responses were collected to gain insight into the state of early career communities, the impacts of federal grant delays on early career astrobiologists, and the future of astrobiology. Results indicated that a majority of surveyed ECRs are concerned about their future careers (>90%). Funding delays have affected 89% of ECRs' future career paths, with open responses indicating that the lack of available funding, resources, and employment opportunities limit their ability to see a future in the field. Most ECRs are unsure about staying in academia, with over 50% of Ph.D. students, postdocs, and Junior scientists 'unsure' or 'not planning to' stay in academia. However, Junior Faculty are largely planning to remain in academia.

## Introduction:

Astrobiologists are faced with unique challenges as they work on the interface of multiple scientific disciplines, from geology and chemistry to planetary science. Early career researchers (ECRs) face unique challenges compared to mid- and late-career researchers, as their future in astrobiology is contingent on available jobs and funding opportunities during the transition from graduate school to permanent employment. In addition to their funding largely being dependent on advisors and PIs, they must also navigate their scientific development, the feasibility of their future career paths, and personal responsibilities. Bridle et al. (2013) stated that ECRs are at a formative stage of their research career and need structured induction to understand and evaluate the opportunities and risks before deciding to make an enduring commitment. Mula et al. (2021) describe ECRs as highly vulnerable figures who face a series of difficulties in their career and struggle to achieve a stable professional position. Federal funding agencies, such as NSF and NASA, influence all academics in these scientific disciplines, including early-, mid-, and late-career scientists. Recognizing the resulting trends from federal funding delays and cancelations in the early-career community can provide vital insight into the future of astrobiology. With federal funding delays and cancelations limiting available opportunities for ECRs during a crucial period in their careers, ensuring the retention of talent and maintaining a robust ECR workforce is essential for the continuation of the field.

These concerns of a loss of highly skilled talent are compounded by long-standing demographic challenges within the federal science workforce. An aging NASA workforce has been repeatedly documented (National Research Council 2006; National Research Council 2007). For example, in 2017 at NASA's Kennedy Space Center (KSC), the average employee age was 47, with 15% eligible for retirement and 34% projected to be eligible within five years; within KSC's Safety and Mission Assurance (S&MA) organization, the average age was 49, with 25% eligible for retirement and 43% within five years (Nguyen et al. 2017). For comparison, the median age of the total U.S. workforce was 41.7 in 2024 (U.S. Bureau of Labor Statistics, 2025). This pattern is reflected across other federal research agencies; across all of NOAA, two-thirds of employees fell within Generation X, Baby Boomer, or older generational groupings (NOAA Office of Inclusion and Civil Rights, 2023). The NASA Office of Inspector General (OIG) reported in 2023 that the average age of NASA civil service employees was 48, with 23% of the federal workforce eligible for retirement. Nearly 40% of NASA's science and engineering workforce fell within the 55-and-over age range, many of whom are eligible for retirement (NASA OIG 2023, NASA's Efforts to Increase Diversity in Its Workforce, IG-23-011). The OIG further noted interrelated challenges, including an aging civil service workforce and a growing shortfall of employees qualified in critical technical areas.

Simultaneously, federal research agencies have experienced significant losses of highly trained scientists. According to employment data analyzed by Science, 10,109 doctoral-trained experts in STEM or health fields exited federal service in a single year representing 14% of the total federal STEM Ph.D. workforce employed at the end of 2024 (Hersher, M. and Mervis, J. (2025)). Across 14 research agencies examined, departures outpaced hires by a ratio of 11 to 1, resulting in a net loss of 4,224 STEM Ph.D.s. Research-intensive agencies were particularly affected: the National Science Foundation (NSF), the Environmental Protection Agency (EPA), Department of Energy (DOE), and the U.S. Forest Service (USFS) all experienced disproportionate losses in their doctoral-level workforce (Hersher, M. and Mervis, J. (2025)). Taken together, an aging workforce and significant attrition of highly trained federal scientists create a structural vulnerability for the scientific enterprise in the United States. For astrobiology, a field that relies heavily on NASA coordination, NSF funding, and interagency collaboration, these trends raise urgent questions about workforce continuity, mentorship pipelines, and institutional stability. As senior scientists retire and federal research capacity contracts, the burden of sustaining the field will increasingly fall on the emerging generation of scientists.

The urgency of understanding and supporting the early career astrobiology workforce has also been explicitly recognized at the agency level. As part of the 2025 NASA Decadal Astrobiology Research and Exploration Strategy (DARES), NASA issued a community Request for Information (RFI) that included "Strengthen Community" as one of its nine requested response topics (Topic 5). Subsequent analysis by the NASA DARES Task Force 1 identified nine focus areas, informed by the RFI Findings Workshop held in May 2025, as well as existing community guidance documents (e.g., the 2015 NASA Astrobiology Strategy and National Academies decadal surveys). Notably, as mentioned in the preliminary findings presented by leadership at AbSciCon on May 18th Focus Area 8, Workforce and Early Career Support, emphasized that addressing fundamental questions in astrobiology cannot occur without a thriving workforce; that sustained support for ECRs is critical to mission success; and that such support must extend beyond research funding to include career development. These priorities underscore the importance of empirically assessing the current experiences and perceptions of early career astrobiologists, particularly in a funding and institutional landscape that is increasingly uncertain.

Research Coordination Networks (RCNs) foster collaborative environment for selected proposal PIs that is directed towards the advancement of the RCN's core research areas. They comprise five networks, each geared towards specific scientific areas relevant to Astrobiology; The Prebiotic Chemistry and Early Earth Environments Consortium (PCE3) is focused on the origins of life, while LIFE: Early Cells to Multicellularity focuses on the evolution of life. Research into the distribution of life is organized into three RCNs: the Network for Ocean Worlds (NOW), the Network for Life Detection (NFoLD), and the Nexus for Exoplanet System Science (NExSS). These RCNs are managed by NASA and operated by community members, and serve as hubs for the astrobiological community. Additionally, the Scientific Society for Astrobiology (SSA) acts as an independent and professional organization to advocate for the interests of astrobiology, and to provide a mechanism for connection and collaboration between all astrobiologists. Many of these groups contain early-career subgroups, focused on supporting early-career astrobiologists within their purview. With the RCNs and the SSA in a prime position to reach the early career community, they are an accessible starting point for assessing the state of ECR astrobiologists. Obtaining empirical data on the state of ECR astrobiologists within these organizations can provide insight into the state of the ECR community, and provide direction on best practices for support.

Considering the vulnerable nature of ECRs and current uncertain funding conditions, the Network for Ocean Worlds (NOW) RCN's early career group: Future Leaders of Ocean Worlds (FLOW), in collaboration with other early career groups, conducted a survey of ECR astrobiologists to assess the state of the field. These results can provide vital insight into the current trends within early career astrobiology, as well as provide insight into the future of astrobiology as a field. We present here an analysis of survey results collected from November 2025–February 2026, as well as potential solutions and support systems submitted in open responses that would increase confidence and retention of early career astrobiologists.

**Survey Description:**

In 2025, a survey was conducted of self-identified early-career astrobiologists, with 'early career' defined as Ph.D. students, postdocs, and Junior Faculty/Scientists (defined as < 10 years post-terminal degree). The survey consisted of 3 parts: demographic questions pertinent to astrobiology, questions about respondents' attitudes toward the current state of the field, and a free-response section in which respondents could report their specific concerns and suggest potential supportive measures. Questions regarding the state of the field were selected to gain insight into the immediate effects of the current institutional uncertainty (ex. funding cancelations/delays, job security) and how this uncertainty translates into ECR's attitude towards a future in astrobiological research.

Trends from the open responses will be reported; however, individual responses will not be shown to protect respondent privacy.

Demographic questions were asked to identify the respondents' career stage and area of specialty. To encourage participation, demographic data was restricted to career-relevant questions. Distribution of the survey was directed at early career entities well known in the astrobiology community, such as the RCN and SSA mailing lists. Respondents were allowed to select 'other' as a career stage. Based on open responses, some members in the 'other' career stage included undergraduate students and those in industry. In a four-month period from November 2025–February 2026, 165 responses were submitted: 66 Ph.D. Students, 26 postdocs, 9 Junior scientists, 7 Junior Faculty, and 35 'Other'. Ph.D. students were the most represented group, and Junior Faculty were the least represented group. During review of the submissions, 22 'other' responses were removed from analysis as they were not relevant and/or spam. Data presented here represent 142 respondents overall, though not every question was compulsory. A breakdown of the respondent career stage demographics is shown in Figure 1.

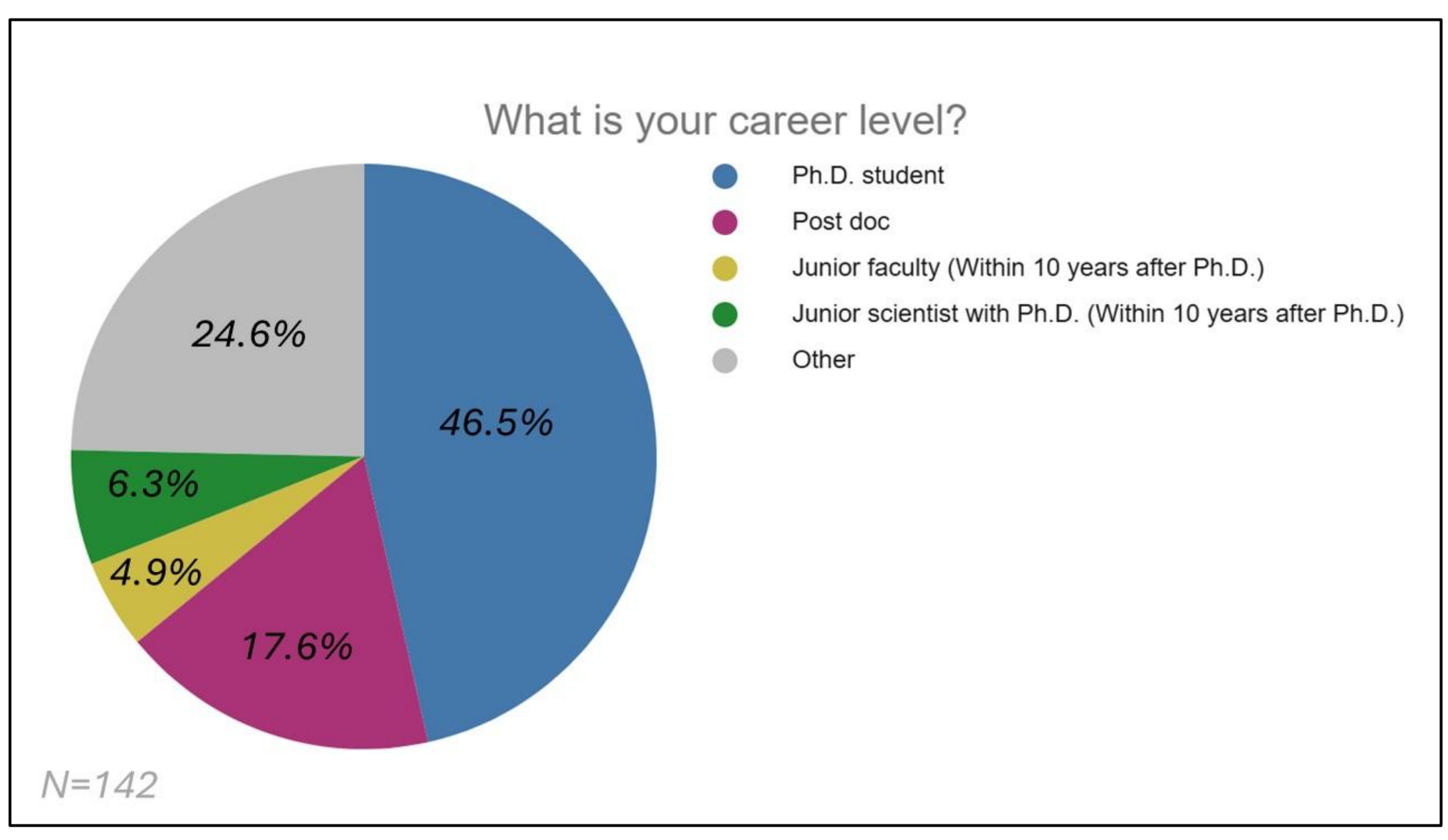

**Figure 1:** Survey respondents' breakdown by career stage.

Respondents were asked to identify their scientific specialty from 'Origins and Evolutions of Life', 'Life in Extreme Environments', 'Habitability and Life in the Solar Systems', 'Exoplanets; Detection Habitability and Biosignature Detection', 'Space Explorations', and 'Other'. While survey individuals span a range of scientific disciplines (Figure 2), the respondents were able to select multiple specialties.

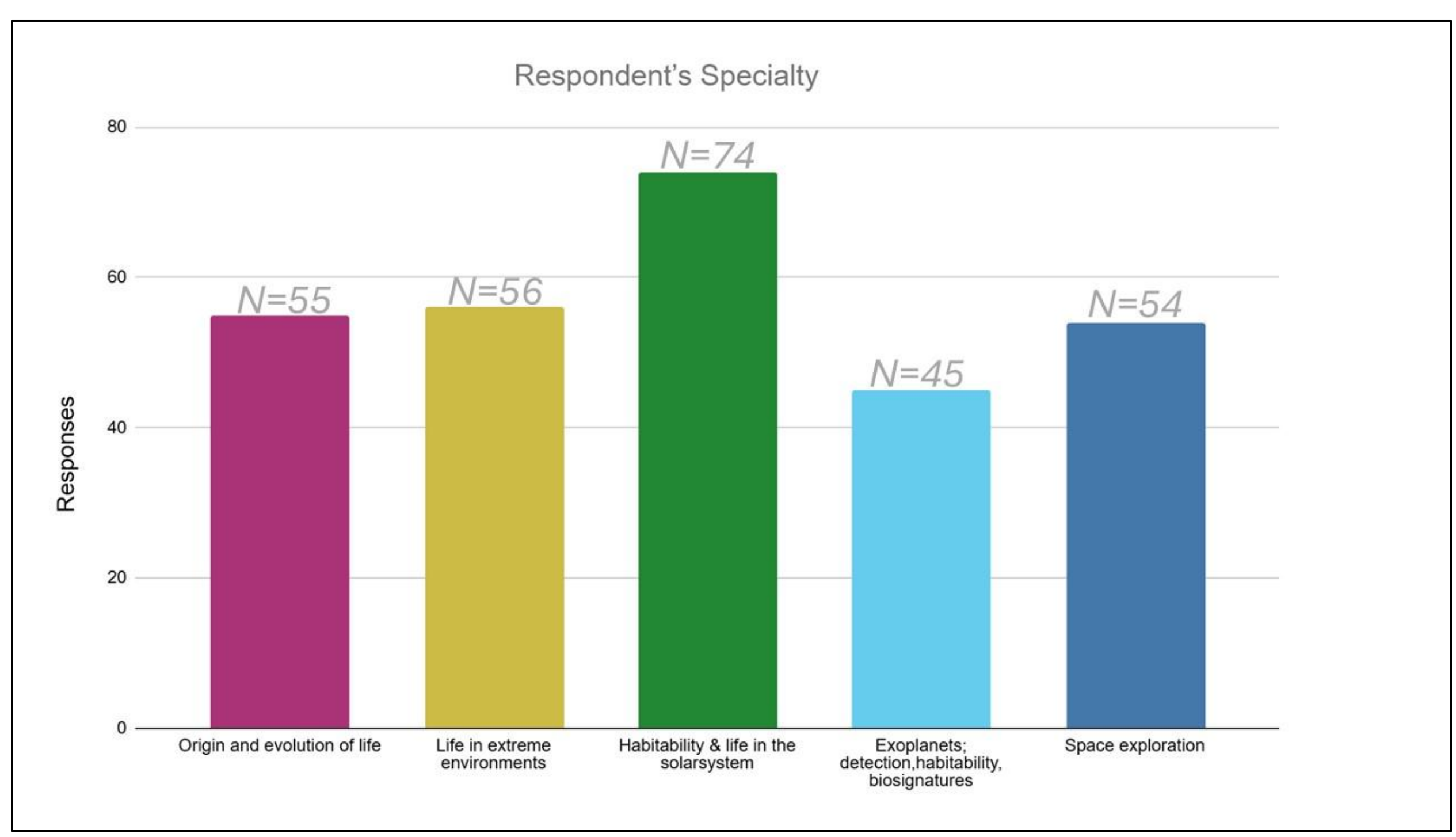

**Figure 2:** Summary of survey respondents' self-identified specialties, with “Habitability and Life in the Solar System” being the most represented, and "Exoplanet detection, habitability, and biosignatures” being the least represented. Note that respondents could select multiple specialties.

We note our survey results are limited by a small sample size and an over-representation of Ph.D. students. Additionally, the survey targets self-identified astrobiologists and scientific specialties. However, as previous ECR surveys have used similar sample sizes (ex. 80 respondents Davey et al. 2024), low response rates can be informative for ECR surveys. Given the niche nature of astrobiology and the natural attrition expected between ascending career stages, the results presented here can still provide valuable feedback on the ECR community,

**Results**

After demographic data was collected, questions on respondents' attitudes toward the current state of the field were asked to gauge the impact of federal funding delays and cancelations on the early career community. Two questions were posed to gauge the impact of funding delays in 2025 by federal agencies on ECR astrobiologists current career: “Did delayed ROSES this year impact your career?” and “Is delayed or canceled funding from any government agencies impacting your current career?”. Responses indicate that 41.7% of ECRs surveyed were not impacted by the ROSES delay (Figure 3); however, when the question is expanded to include “delayed or canceled funding from any government agencies,” 68.6% of ECRs surveyed report being impacted (Figure 4), an increase of 26.9%. Junior scientists surveyed were the most likely to be negatively impacted, with 100% of Junior scientists ‘somewhat’ or ‘strongly’ impacted by funding delays or cancelations from any government agency (Figure 4). However, all ECR groups surveyed responded > 50% that they were ‘somewhat’ or ‘strongly’ impacted by funding delays from government agencies. Open responses indicate that this could be related to the reduction in jobs and opportunities stemming from budget constraints.

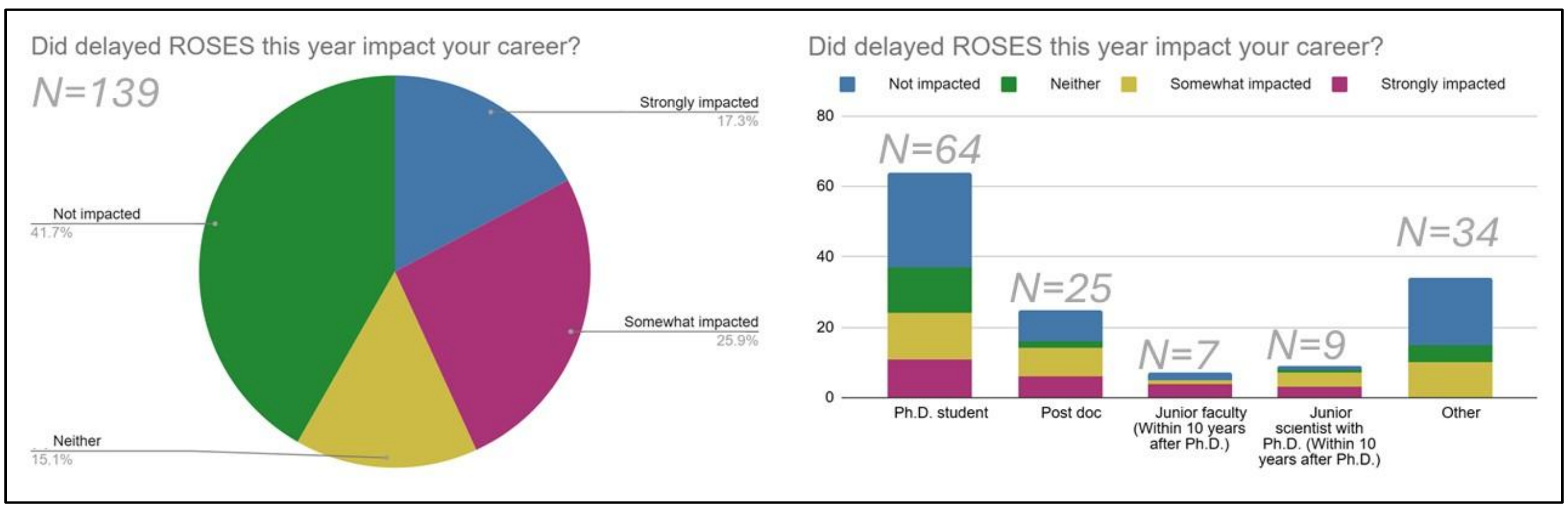

**Figure 3:** Responses to questions regarding the impact on ECRs being impacted by ROSES funding delays. ROSES delay impacts to current ECR careers were reportedly split, with respondents reporting either not impacted (41.7%) or ‘neither’ impacted or not impacted (15.1 %) totaling 56.8%. However, 43.2% of respondents reported ROSES delays ‘strongly impacted’ (17.3%) or ‘somewhat impacted’ (25.9%) their career. Junior Faculty and Junior scientist reported the highest demographics of being impacted (71% and 77% respectively).

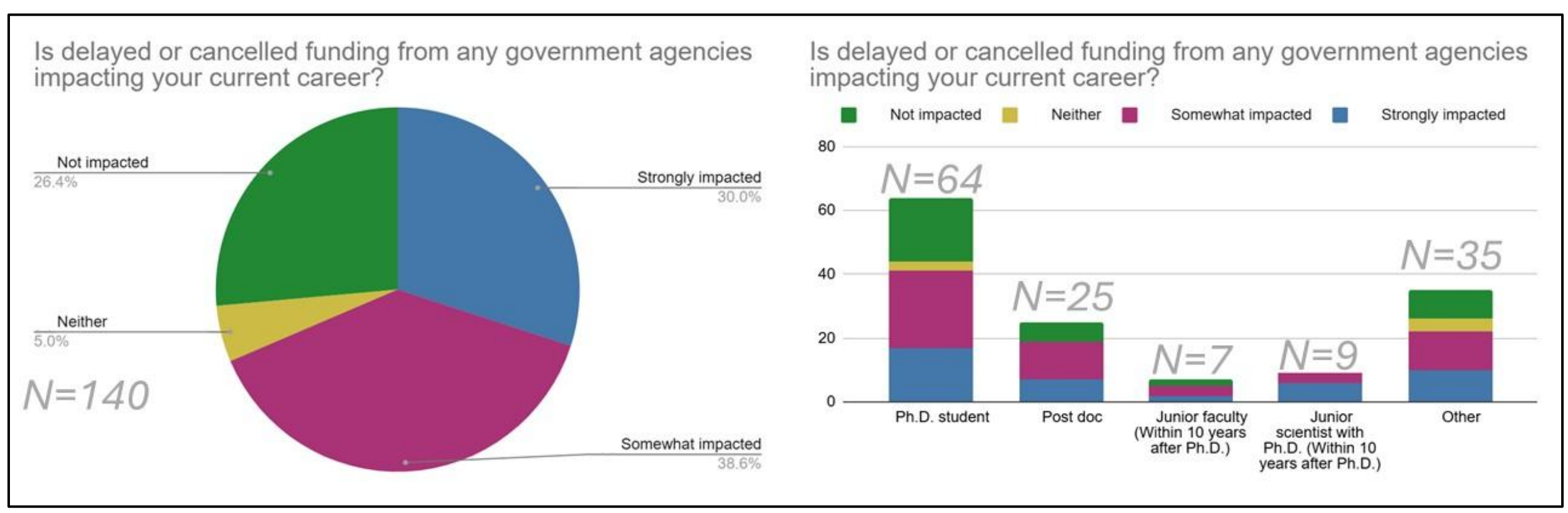

**Figure 4:** Responses to questions regarding the impact on ECRs being impacted by government funding delays, and responses broken down by career level. Every group indicated > 50% of being impacted (PhD student 64%, Postdoc 76%, Junior Faculty 71%, Junior scientist 100%, and Other 62%).

Questions regarding ECR’s view on their future careers were included in the survey. When asked, “Is delayed or canceled funding from any government agency impacting your future career?” 89.3% of ECRs surveyed responded that their future career was impacted by funding delays and cancelations (Figure 5), representing a 22.4% percentage point increase in future concerns relative to current conditions. Additionally, responses to the question, “Are you currently concerned about your science career?”, indicated that there is an overwhelming concern (93.6%) about ECR’s career in the scientific field (Figure 6). Finally, when asked if respondents are planning to stay in academia, more than 50% of Ph.D. students, postdocs, and Junior scientists surveyed responded "no" or that they were "not sure". The only group that deviated from this trend was the Junior Faculty, who largely (85%) responded "yes" they planned to remain in academia (Figure 7).

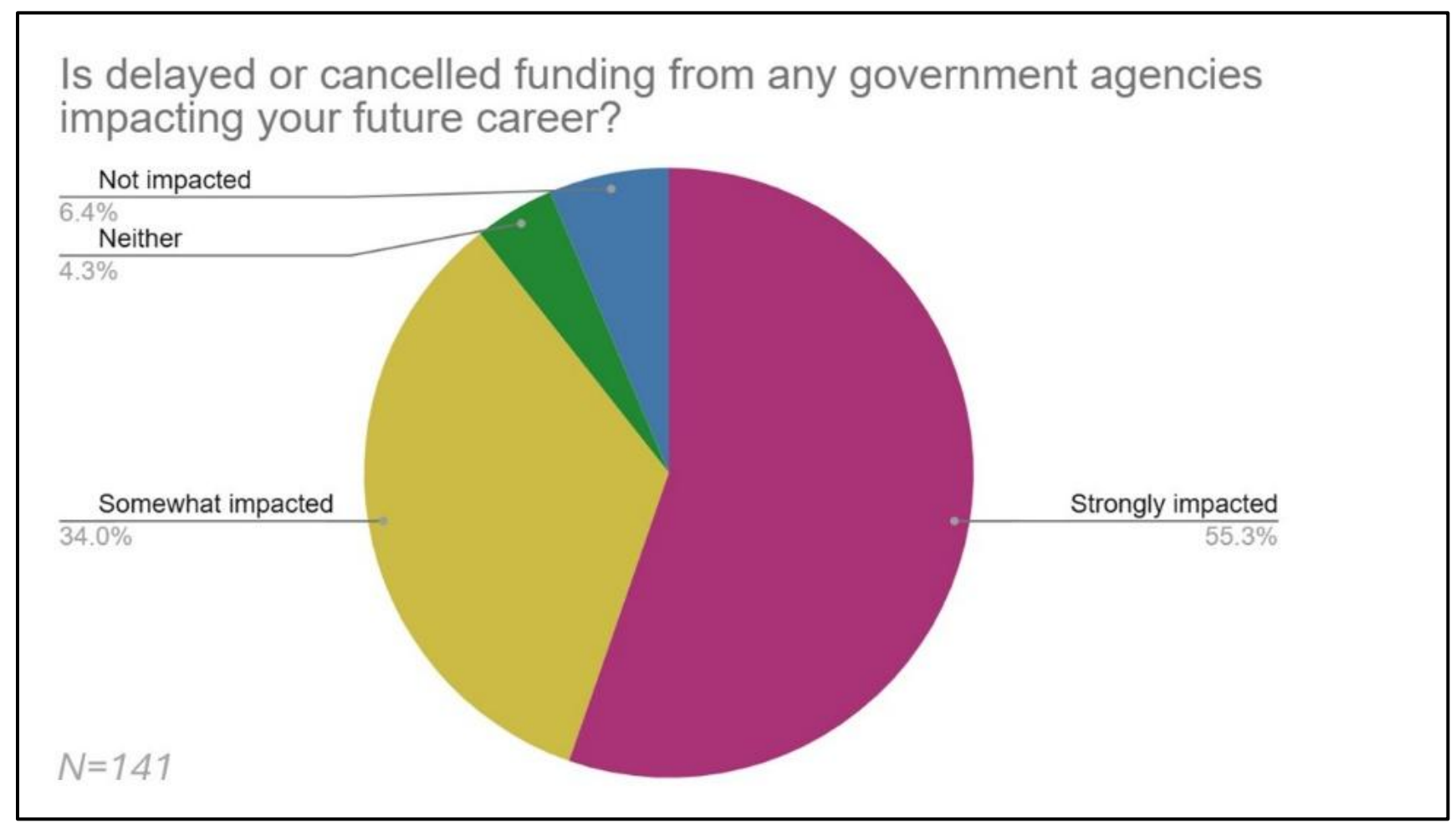

**Figure 5:** When asked if delayed funding had affected ECR's future careers, respondents largely (89%) responded they had been "somewhat impacted" and "strongly impacted". Open responses indicate that these answers are sourced not only from the lack of current opportunities to continue in the field, but also a perception that the field lacks stability and future job prospects.

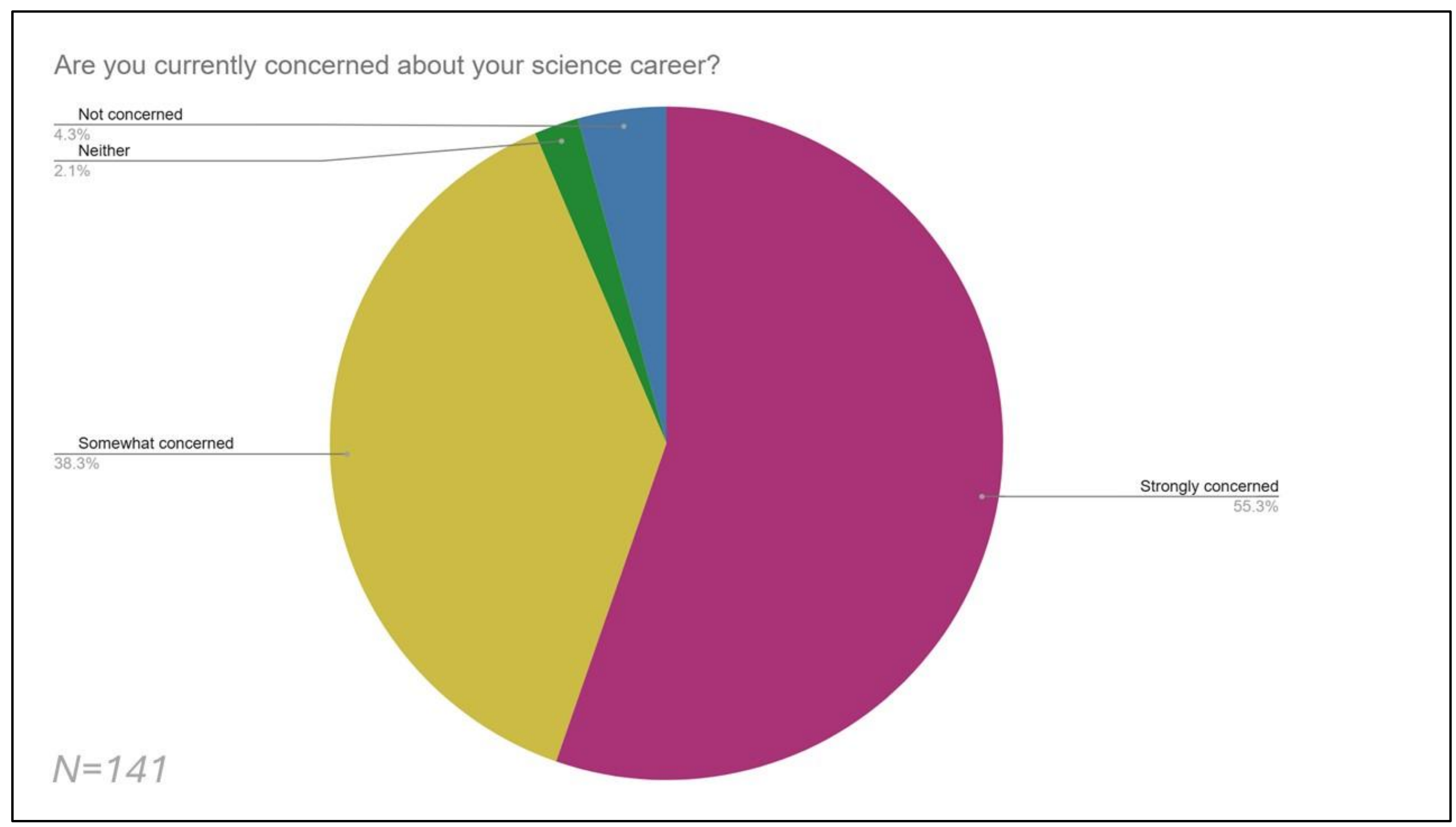

**Figure 6:** Responses to the question, "Are you currently concerned about your science career?", indicated that an overwhelming majority of ECRs are concerned (93.6%), with 55.3% reporting they are 'strongly concerned' and 38.3% reporting they are 'somewhat concerned' about their science career.

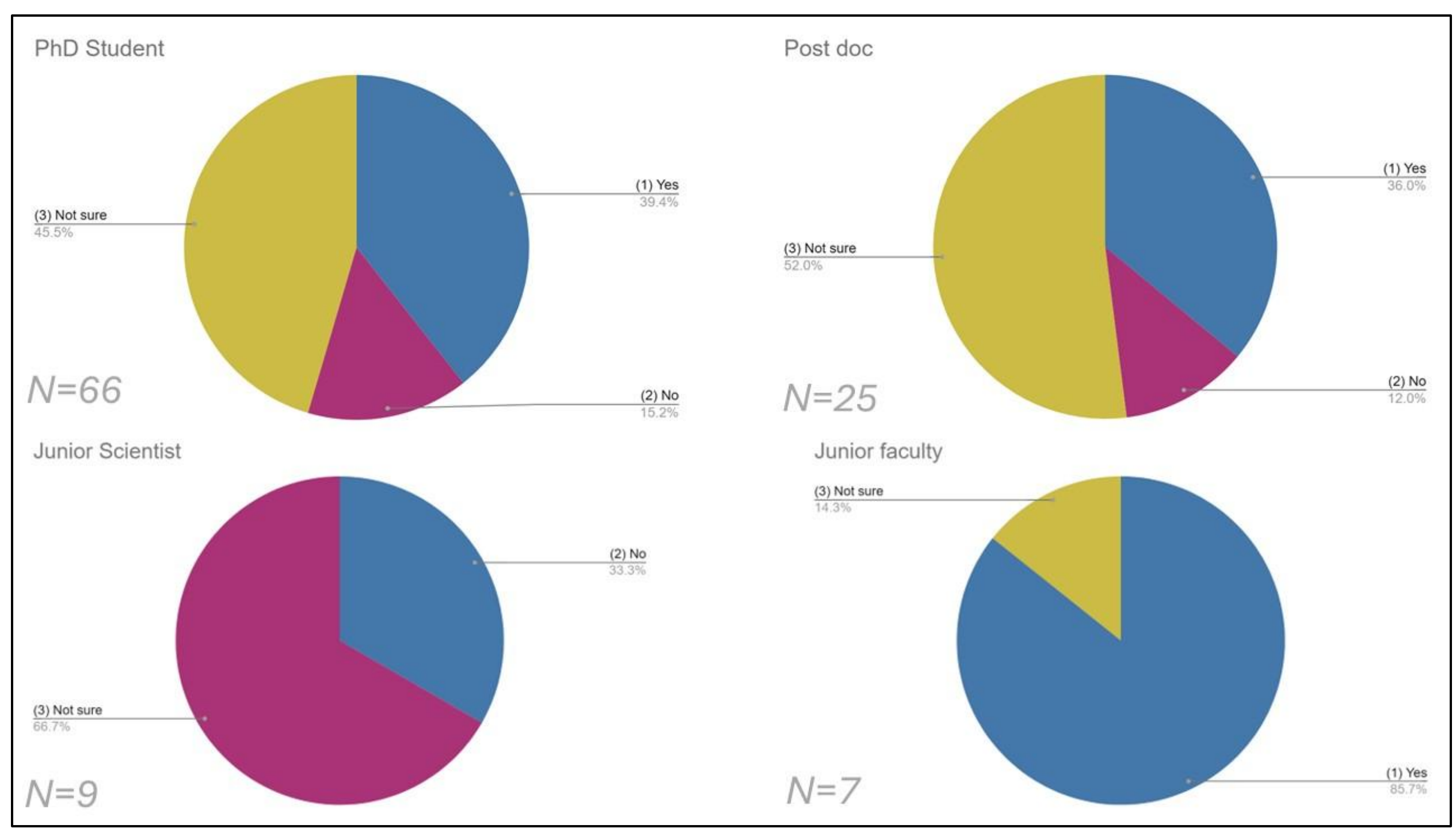

**Figure 7:** Responses to the Survey question "Are you planning to stay in academia?". When surveyed, > 50% of Ph.D. students, postdocs, and Junior scientists surveyed responded 'no' or that they were 'not sure' if they were planning to stay in academia.

## Discussion and Submitted Solutions

Overwhelmingly, the results presented here indicate that the ECR community is being negatively impacted by the current academic landscape. Results from questions gauging current ECR trends indicate that a majority of ECRs are actively impacted by federal funding delays (though this is not attributed to NASA specifically). Questions regarding future career direction capture the apprehension and uncertainty of the future of the field. Even with the comparatively smaller impact of federal funding delays to students' current careers, the stability and prospects for ECR's future in science have undoubtedly been altered, as observed in the discrepancies observed in responses inquiring about ECRs current and future careers. The 22.4% percentage point increase from delays/cancelation of federal grants impact on current to future careers provides insight into the lack of opportunity (real or perceived) many are observing in the field. It is therefore unsurprising that many ECRs are concerned about their scientific careers (93.6%), and that a majority of ECRs are not confident about staying in the field of astrobiology. It is clear that the state of the scientific field is discouraging ECRs to remain in the field. A potential loss of this talent could escalate the already precarious need for a robust scientific workforce.

There are several studies investigating the well-being of ECRs in academia (Cilli et al., 2023; Guo and Liu, 2025). Cilli et al. (2023) surveyed 134 ECRs and reported negative psychological dimensions towards long-term goals and motivation due to high anxiety and stress, especially among Ph.D. students. Guo and Liu (2025) focused on Junior faculty for their survey and reported that high anxiety is impacting their performance and well-being. These studies support that ECRs are already in a vulnerable status due to the current academic systems. On top of these challenges due to the unique nature of ECRs, our study supported that recent delays and cuts for NSF and NASA are highly impacting ECRs. When assessing impacts from federal funding delays, our results showed ECRs' current careers might be impacted (68.6%); however, their future careers will be mostly impacted (89%). This increase in ECR's impact from their current career to their future career highlights that systematic changes need to be established in a timely manner to keep future leaders in astrobiology. Open responses indicate that there is a perception that funding and career options are not believed to be available during ECR transitions. Additionally, open responses indicate a significant disconnect between these articulated community priorities identified by the NASA TF1 and the lived experiences of ECR astrobiologists, suggesting that key workforce needs identified by NASA remain inadequately addressed.

Funding and career development challenges are only compounded by the interdisciplinarity inherent to the field of astrobiology. Between difficulties in peer review (Pautasso & Pautasso, 2010), the departmental structure of universities (Zheng et al., 2025), and funding barriers (Bromham et al., 2016), literature documenting the challenges for ECRs in interdisciplinary fields is growing. Applying biodiversity metrics, an analysis of over 18,000 proposals to the Australian Research Council's Discovery Programme found that proposals with a higher degree of interdisciplinarity had a significantly lower probability of being funded, even when the number of collaborators, primary research field, and type of institution were accounted for (Bromham et al., 2016). A similar analysis of 154,000 researchers who received a Ph.D. between 1970 and 2013 characterized the disciplinary composition of references and found that authors of the most interdisciplinary publications stopped publishing much earlier in their careers than those with less interdisciplinary focus, quantifying real impediments to career development in interdisciplinary fields (Berkes et al., 2024). At the same time, interdisciplinary research can be associated with higher scientific impact, suggesting a structural mismatch (Leahey et al., 2017). Taken together with the uncertainty highlighted in our survey results, these findings underscore the need for targeted support structures for ECRs in astrobiology that explicitly account for the interdisciplinary nature of the field.

In our survey, participants also had an opportunity to answer open questions such as what makes ECRs feel supported, and provide a place to reflect on their concerns and articulate solutions. These open answers can be organized into 4 categories: (1) More funding opportunities exclusively for ECRs, (2) more stable job opportunities, (3) supportive mentor systems, and (4) mental health support and balanced workloads. The responses from ECRs in our survey are relatively consistent with key findings from the RFI synthesis and submitted white papers (ex. Spiers et al. 2025), which identified a range of priorities, including continuing the role of RCNs, increasing funding support through ROSES, incentivizing mentorship, strengthening intra- and interagency collaboration, and investing in ECR bridges. This alignment between ECR needs and emerging NASA-DARES priorities highlights a critical opportunity for action, requiring coordinated support from government agencies and academic institutions to strengthen ECR communities. Synthesizing previous strategy documents, DARES findings, and these survey results, support for the ECR astrobiology community can be summarized by the following actions:

- More funding opportunities – mentioned by 19 respondents
- Mentorship - mentioned by 8 respondents

- Information on international opportunities - mentioned by 4 respondents
- Continued support for ECR involvement and training for missions (Spiers et al. 2025)
- Continued support of the NASA Astrobiology Postdoctoral Program (NPP)
- Provide Astrobiology Graduate fellowships that are explicitly interdisciplinary with 2+ thesis supervisors from different disciplines. (Spiers et al. 2025)
- Incentivize ECR funding and mentorship within data analysis programs (DAP) or similar programs (Spiers et al. 2025)
- Continued support of travel funds such as the Lewis and Clark Award and the NASA Early Career Collaborator Award. These awards enable collaboration opportunities for ECRs across institutional boundaries, enabling research that may not have otherwise been feasible.
- Continued support of AbSciCon and AbGradCon with the addition of initiatives to build peer networks in-person. (Spiers et al. 2025)

Based on these survey results, we encourage NASA and other federal agencies to continue support for ECRs through initiatives like the Future Investigators in NASA Earth and Space Science and Technology (FINESST) Program, NASA Postdoctoral Program (NPP), and similar programs. Additional programs and support might be needed to keep talent in the U.S. astrobiology community. Detailed recommendations and resources for supporting ECRs in astrobiology can be found in Davey et al. (2024), Spiers et al. (2025), and on the NASA Astrobiology website.

**Author contributions:**

K.A.D.: Conceptualization (co-lead); Writing—original draft (lead); Formal analysis (lead); Writing—review and editing (equal). G.R. Writing—review and editing (equal);

Conceptualization (supporting); P.E.J.: Writing—review and editing (equal); Conceptualization (supporting); P.J.M.: Writing—review and editing (supporting); I.M. Writing—review and editing (supporting); E.M.S.: Conceptualization (co-lead); Writing—original draft (lead)

**Statements and declarations:**

The authors declare that no relevant or material financial interests that relate to the research described in this paper.

**Declaration of conflicting interest:**

The author(s) declared no potential conflicts of interest with respect to the research, authorship, and/or publication of this article.

**Funding statement:**

The author(s) received no financial support for the research, authorship, and/or publication of this article.

**Ethics Statement:**

This voluntary survey was conducted independently by the Future Leaders of Ocean Worlds. Survey responses were anonymous, with no personal demographic or identifying information recorded. Results are reported in aggregate.